\documentstyle[12pt,epsf]{article}
\newlength{\abstractwidth}
\renewcommand{\thefootnote}{\fnsymbol{footnote}}
\renewcommand{\thanks}[1]{\footnote{#1}} 
\newcommand{\starttext}{
\setcounter{footnote}{0}
\renewcommand{\thefootnote}{\arabic{footnote}}}
\renewcommand{\theequation}{\thesection.\arabic{equation}}
\newcommand{\be}{\begin{equation}}
\newcommand{\bea}{\begin{eqnarray}}
\newcommand{\eea}{\end{eqnarray}}
\newcommand{\beq}{\begin{equation}}
\newcommand{\ee}{\end{equation}}
\newcommand{\eeq}{\end{equation}}

\newcommand{\m}{\mu}

\def\ba{\begin{eqnarray}}
\def\ea{\end{eqnarray}}

\def\12{{1 \over 2}}
\def\32{{3 \over 2}}
\def\72{{7 \over 2}}
\def\92{{9 \over 2}}

\def\nc{non--commutative}
\def\ny{non--commutativity}

\def\dz{$D0$-brane}
\def\dzs{$D0$-branes}
\def\dt{$D2$-brane}

\def\ds{$D6$-brane}
\def\dss{$D6$-branes}

\def\cs{Chern Simons}
\def\qp{quasiparticle}
\def\qhs{Quantum Hall Soliton}

\begin{document}
\renewcommand{\theequation}{\thesection.\arabic{equation}}
\begin{titlepage}
\bigskip
\rightline{SLAC-PUB-8657} \rightline{SU-ITP 00-25}
\rightline{hep-th/0010105}

\bigskip\bigskip\bigskip\bigskip

\centerline{\Large \bf {How Bob Laughlin Tamed the Giant
Graviton}}
\centerline{\Large\bf{ from Taub-NUT space}}

\bigskip\bigskip
\bigskip\bigskip

\centerline{\it B.~A. Bernevig $^1$, J. Brodie $^2$, L. Susskind
$^1$ and N. Toumbas $^1$ }
\medskip
\medskip
 \centerline{$^1$ Department of Physics} \centerline{
Stanford University} \centerline{Stanford, CA 94305-4060}
\medskip
\medskip
\centerline{$^2$ Stanford
Linear Accelerator Center} \centerline{Stanford
University}\centerline{Stanford, CA 94309-4349}
\bigskip\bigskip
\begin{abstract}
In this paper we show how two dimensional electron systems can be
modeled by strings interacting with $D$-branes. The dualities of
string theory allow several descriptions of the system. These
include  descriptions in terms of solitons  in the near horizon
$D6$-brane theory, non-commutative gauge theory on a $D2$-brane,
the Matrix Theory of $D0$-branes and finally as a giant graviton
in M-theory. The soliton can be described as a $D2$-brane with an
incompressible fluid of $D0$-branes and charged string-ends moving
on it. Including an $NS5$ brane in the system allows for the
existence of an edge with the characteristic massless chiral edge
states of the Quantum Hall system.
\medskip
\noindent
\end{abstract}

\end{titlepage}
\starttext \baselineskip=18pt \setcounter{footnote}{0}

\setcounter{equation}{0}
\section{Introduction  }
The dualities of string theory have provided powerful tools for
the study of strongly coupled quantum field theories. The most
surprising of these dualities involves field theory on one side of
the duality and gravitation on the other. For example, Matrix
Theory ~\cite{BFSS} relates Super Yang Mills theory on various
tori to compactifications of 11 dimensional supergravity.
Similarly the ADS/CFT duality
~\cite{Maldacena}\cite{GKP}\cite{Witten1}\cite{AGMOO} relates
large $N$ gauge theories to supergravity in an Anti deSitter
background. The result is that many problems of quantum field
theory such as confinement
~\cite{Witten2}\cite{Brandhuber}\cite{Girardello}\cite{KS} and
finite temperature ~\cite{Witten2}\cite{Rey} behavior are solved
by finding classical solutions of the appropriate gravity
equations. These solutions include black holes, gravitational
waves and naked singularities
~\cite{Johnson}\cite{PS}\cite{Gubser}.

In view of all these, one may hope that similar correspondences
exist involving interesting condensed matter systems. The purpose
of this paper is to demonstrate a correspondence between certain
solitons in the near horizon geometry of a $D6$-brane and the
Quantum Hall System ~\cite{Girvin}\cite{susskind} -- charged
particles moving on a two dimensional surface in the presence of a
strong magnetic field. Additional dualities allow a descriptions
in terms of \dz \ matrix quantum mechanics ~\cite{BFSS} and giant
gravitons in M-theory.

\setcounter{equation}{0}
\section{The Brane Setup }

We work in uncompactified $IIA$ string theory. Let us begin with a
coincident stack of $K$ $D6$-branes whose worldvolume is oriented
along the directions $(t, Y^a)$, where $a = 4,...,9$. The three
remaining directions we call $X^i$, $i = 1,2,3$. The $D6$-brane is
located at $X^i = 0$.

Now let us add a spherical $D2$-brane wrapped on the sphere $S_2$;
\be
\sum_{i=1}^3 {(X^i)}^2 = r^2. \ee For the moment let us ignore the
stability of this configuration. We would like to show that
consistency requires the presence of $K$ fundamental strings
connecting the $D6$-branes and the $D2$-brane. To see this, we
recall that the $D6$-brane acts as a magnetic source for the
Ramond-Ramond gauge field $C_{1}$ which couples electrically to
$D0$-branes. In other words the $D6$-brane acts as a magnetic
monopole situated at $X^{i}=0$. Let $\tilde{H}_2$ denote the
$2$-form field strength of $C_{1}$. The flux through the sphere is
then given by
\be
\int_{S_2}\tilde{H}_2= 2\pi K\mu_6, \ee where $\mu_6$ denotes the
elementary $D6$-brane charge. Note also that Dirac's quantization
condition requires that
\be
\mu_6 \mu_0 = 1, \ee where $\mu_0$ is the $D0$--brane charge.
Evidently, the field strength $\tilde{H}_2$ is given by
\be
\vec{H}(\vec{r})= {K\mu_6 \hat{r} \over 2 r^2}, \ee where
$H_i={\epsilon_i}^{jk}\tilde{H}_{jk}/2$.

Next recall that $\tilde{H}_2$ is coupled to the $D2$-brane
world-volume gauge field $A_{\mu}$ through the coupling
\be
 {\mu_2 \over 2}\int_{S_2}
\epsilon^{\mu\nu\lambda}(2 \pi
\alpha')A_{\mu}\tilde{H}_{\nu\lambda} = \int_{S_2} J^0A_0 \ee with
\be
J^{0}={\mu_2} (2\pi\alpha') |\vec{H}(\vec{r})|= {\mu_2 \mu_6 K
(2\pi \alpha') \over 2 r^2} . \ee The expression above corresponds
to a background charge density on the $D2$-brane with total charge
\be
Q=  2\pi K \mu_2 \mu_6 (2\pi \alpha'). \ee Since branes are BPS
objects the ratio of their charges is equal to the ratio of their
tensions. Thus $\mu_2 = \mu_0 (T_2 /T_0)$. Then using eq. (2.3),
we obtain
\be
Q= 2\pi K (2\pi \alpha'){T_2 \over T_0} = K. \ee This background
charge must be cancelled since the total charge on a compact space
must vanish. Thus we must add $K$ strings stretched between the
$D6$ and the $D2$ branes.

This result is closely related to the Hanany-Witten effect
~\cite{HW}. Begin with the $D2$-brane far from the $D6$-branes and
not surrounding them. Now move the membrane towards the six
branes. As the $D6$-branes pass through the $D2$-brane, the
Hanany-Witten effect adheres the strings. Later we will give a
Matrix Theory argument for the same result.

\begin{figure}
\centerline{\epsfysize=2.00truein \epsfbox{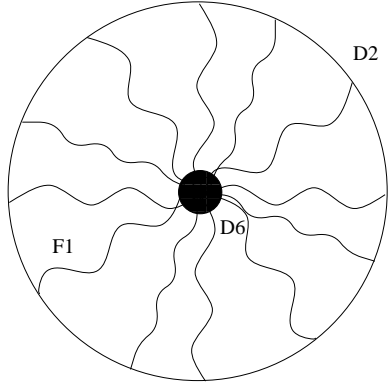}} \vskip
-0.4 cm
 \caption{Stable spherical D2-brane with $N$ units of
magnetic flux surrounding K D6-branes. K fundamental strings
stretch from the D2-brane to the D6-brane. There is a uniform
density of negative charge on the sphere due to the field
generated by the D6-brane.} \label{diagrams}
\end{figure}

\setcounter{equation}0
\section{Balancing The System}
The system as described is not stable. The tension of the
$D2$-brane and the $K$ strings will cause the spherical $D2$-brane
to collapse. To counteract this let us add $N$ $D0$-branes
dissolved in the $D2$-brane. It is well known that $D6$-branes and
$D0$-branes repel one another. The dissolved \dzs \ give rise to a
magnetic flux on the $D2$-brane world volume. The integrated flux
is just the $N$ units of \dz \ charge. As we will see the
repulsion can stabilize the radius of the $D2$-brane. The
resulting object we will call a \qhs.

We will work in the approximation that the $D2-D0$ system is a
test probe in the $D6$-brane geometry. In other words we ignore
the backreaction of the $D2-D0$ system on the geometry. The
backreaction will alter the precise details but we do not expect
it to change the scalings that we find.

The string frame metric of the $K$ $D6$-branes is given by
\be
ds_6^2 = h(r)^{-{1 \over 2}}dt^2 - h(r)^{-{1 \over 2}}dy^ady^a -
h(r)^{1 \over 2}dr^2 - h(r)^{1 \over 2}r^2d\Omega_2^2 \ee with
$h(r)$ given by
\be
h(r)= 1 + {Kg_s l_s \over 2r}, \ee where $g_s$ and $l_s$ are the
string coupling constant and length scale. The background dilaton
field is given by
\be
g_s^2 e^{2\Phi}=g_s^2h(r)^{-{3 \over 2}}. \ee We will choose
parameters so that the $D2$-brane is well within the near horizon
region in which we can set
\be
h(r)= {Kg_sl_s \over 2r}. \ee

It will be convenient to rescale the co-ordinates
\be
\tilde{y}=\left({Kg_s \over 2}\right)^{1\over 3}y, \,\
t=\left({Kg_s \over 2}\right)^{1 \over 3}\tau, \,\ r=\left({Kg_s
\over 2}\right)^{-{1 \over 3}}\rho \ee so that the metric becomes
\be
ds^2 = \sqrt{\rho \over l_s}(d\tau^2 -
d\tilde{y}^ad\tilde{y}^a)-\sqrt{l_s \over \rho}(d\rho^2 +
\rho^2d\Omega_2^2).\ee The background dilaton becomes then
\be
g_s^2 e^{2\Phi} = {4 \over K^2} \left({\rho \over l_s}\right)^{3
\over 2}. \ee

Now consider the  $D2$--brane wrapped on the $2$-sphere with $N$
units of \dz \ charge or equivalently, $N$ units of magnetic flux.
As in the previous section, we must add $K$ $6-2$ strings which we
orient along the radial direction. The action for the $D2$--brane
in the background geometry is given as usual by the Dirac Born
Infeld (DBI) action. From the DBI action of the brane plus the
action of the strings we can obtain a potential for the radial
mode $\rho$.

We choose worldvolume co--ordinates such that
\be
\xi^0=\tau,\,\ \xi^1=\theta,\,\ \xi^2=\phi. \ee Dropping time
derivatives, the induced metric on the brane $G_{ab}$ becomes
\be
G_{00}= \sqrt{\rho \over l_s}, \,\ G_{11}=- \sqrt{\rho^3l_s}, \,\
G_{22}=-\sqrt{\rho^3l_s}\sin^2\theta. \ee In addition, there are
$N$ units of flux on the brane:
\be
\int_{S_2}F=2\pi N. \ee Thus the background field on the brane is
given by
\be
F_{12}= {N \over 2}\sin\theta. \ee This corresponds to a constant
field strength perpendicular to the 2-sphere.

The DBI Lagrangian for the brane is then given by
\be
L_{D2}=-{1 \over 4\pi^2 g_s l_s^3}\int d\theta d\phi
e^{-\Phi}det[G_{ab}+2\pi l_s^2 F_{ab}]^{1 \over 2}=-{K\rho \over
2\pi l_s^2}\sqrt{1 + {\pi^2 N^2 l_s^3 \over  \rho^3}}. \ee The
contribution of the $K$ strings is given by
\be
L_{Strings}= -{K \over 2\pi l_s^2}\rho. \ee Therefore, the
potential for $\rho$ becomes
\be
V(\rho)={K\rho \over 2\pi l_s^2} \left(\sqrt{1 + {\pi^2 N^2 l_s^3
\over  \rho^3}} + 1\right). \ee The potential has a minimum at
\be
\rho_* = {({\pi N})^{2 \over 3} \over 2}l_s \ee for all $N$ and
$K$. Thus the brane can stabilize at this co--ordinate distance.

We require that our brane lives in the near horizon region.
Therefore, we must have
\be
h(\rho_*) > 1 \ee or that
\be
g_s > {\sqrt{N} \over K} \ee at infinity. For fixed $K/N$ and any
value of $g_s$ this will be satisfied for large enough $N$.

The proper area of the stable membrane is given by
\be
A= 4\pi \sqrt{{\rho_*}^3l_s} =\sqrt{2}\pi^2 N l_s^2. \ee The fact
that the \dz \ density is universal in string units is noteworthy.
It means that the \dz \ system is behaving like an incompressible
fluid. This also implies that the magnetic field and the magnetic
length is fixed.

Let us next consider the gauge coupling of the theory on the
$D2$-brane. The theory is an abelian gauge theory with coupling
constant given by
\be
g_{YM}^2l_s =g_se^{\Phi}|_{\rho_*} = {2 \over K} \left({\rho_*
\over l_s}\right)^{3 \over 4} = 2^{1 \over 4}{\sqrt{\pi N} \over
K} \ee independent of $g_s$ at infinity.

In the Quantum Hall interpretation of the system $K$ plays the
role of the number of charged particles and $N$ the total magnetic
flux. The ratio $K/N = \nu$ is the filling fraction which we will
want to keep fixed as $N \rightarrow \infty$. We therefore find
\be
g_{YM}^2 \sim {1 \over \nu \sqrt{N} l_s}.\ee

We also note that the curvature of the background geometry at
$\rho_*$ is given by
\be
l_s^2 {\cal{R}} \sim {1 \over N} \ee and so it is weak for large
$N$. Thus we can reliably use the DBI action to study the dynamics
of the system. Finally, the $D0$--brane magnetic field through the
sphere is fixed in string units $|\vec{H}| \sim \mu_6  K/N l_s^2
\sim \mu_6\nu/l_s^2 $.

\setcounter{equation}0
\section{Energy Scales}
In this section we will see that a single energy scale controls
the dynamics of the \qhs . In discussing these energy scales we
will work in units appropriate to a local observer at the
$D2$-brane. In other words, let us once again rescale time so that
proper time at the $D2$-brane is $T$:
\be
dT^2=\sqrt{\rho_* \over l_s}d\tau^2 \ee or
\be
dT = {({\pi N})^{1 \over 6} \over 2^{1 \over 4}} d\tau. \ee The
energy scales we derive refer to the Hamiltonian conjugate to $T$.

{\bf {Quasiparticle Coulomb Energy.}}
Quantum hall fluids are said to be incompressible. By this
it is meant that the system has an energy gap, namely the energy
of a quasiparticle. Later we will discuss the
formation of fractionally charged quasiparticles \cite{Girvin}.
For the moment, we can just regard a quasiparticle as a localized
object with charge $\pm \nu$ and a radius of order the magnetic
length. It has a Coulomb energy of order
\be
E_{Quas} \sim (g_{YM}\nu)^2, \ee which from eq. (3.20) is
\be
E_{Quas} \sim {\nu \over \sqrt{N}l_s}. \ee This is the basic
energy scale of quantum hall excitations against which other
energies should be compared.

{\bf{Long String Excitations.}} If a string of length $L$ is
vibrationally excited, its energy is of order $1/L$. The proper
length of the $6-2$ strings is of order
\be
L \sim {({\rho_*}^{3}l_s)}^{1 \over 4} \ee and so using eq. (3.16)
we find
\be
L \sim \sqrt{N}l_s \ee corresponding to an energy scale
\be
E_{String} \sim {1 \over \sqrt{N}l_s}.\ee Note that this scales
with $N$ in the same way as the quasiparticle energy but it is
typically bigger by a factor $1/\nu$.

{\bf{Cyclotron Frequency.}} The energy required to excite a higher
Landau level is given by the cyclotron frequency
\be
\omega_{Cycl}= {B \over m}, \ee where $B$ is the magnetic field
and $m$ is the mass of a charge. The charges are strings of mass
$L/l_s^2$ and in string units $B \sim 1$. Therefore,
\be
\omega_{Cycl} \sim {1 \over \sqrt{N}l_s}. \ee Once again this
scales like the quasiparticle energy but it is bigger by the
factor $1/\nu$.

{\bf{Field Theory Gap.}} Since the radius of the 2-sphere is $\sim
N^{1/2}l_s$ the energy of the lowest field mode living on the
\dt \ is of order $1/N^{1/2}l_s$. Later we will see that the gauge
field has a mass of the same order of magnitude.

As we have seen the radius of the $2$-sphere is stabilized by the
competing terms in the DBI potential. The spherically symmetric
fluctuations about this equilibrium are described by a massive
scalar field. By expanding the DBI action to quadratic order in
the fluctuations, we find the mass to again be $ \sim
1/N^{1/2}l_s$. Thus we see a single energy scale governing all of
low energy physics on the membrane.

Again it is noteworthy that a single energy scale appears in the
low energy behavior.  By an additional rescaling of time (which we will not
do)
the energy and time scales for the system can all be made to be of
order unity. Unless $\nu \ll 1$, there is no large separation of energy
scales.
Our assumption will be that despite the lack of large scale
separation the quantum hall effect is robust, at least for $\nu$
not too large.

\subsection{$D0$-Brane Emission}
For finite $N,K$ the \qhs \  can not  be absolutely  stable. The
\dt \ carries no net charge. If the \dzs \ escape from the
membrane they will be repelled to infinity, leaving the \dt \ to
collapse and disappear. We will argue that the emission of a \dz \
is a tunneling process with a barrier that becomes infinite as
$N,K $ become large.

The value of the potential (3.14) at the minimum $\rho_*$ is given
by
\be
V(\rho_*)={KN^{2 \over 3} \over \pi^{1 \over 3} l_s}. \ee This
corresponds to a proper $T$-energy given by
\be
V_T(\rho_*)={2^{1 \over 4}K\sqrt{N} \over \sqrt{\pi}l_s}.\ee
Suppose that the system emits a $D0$-brane so that the flux
changes by one unit. Then to leading order in $1/N$ the change in
the energy is given by
\be
V_T(N)-V_T(N-1)={K \over 2^{3 \over 4} \sqrt{\pi N} l_s}.\ee This
is of the same order of magnitude as the mass of a $D0$-brane just
outside the brane but smaller by a factor of $\sqrt{2}$:
\be
M_{D0}={1 \over g_s(\rho_*)l_s}= {K \over 2^{1 \over 4}\sqrt{\pi
N}l_s}. \ee Therefore, we can estimate the binding energy of a
$D0$-brane to be of order
\be
E_{bind} \sim {\nu N^{1 \over 2} \over l_s}. \ee This binding
energy represents the height of the tunneling barrier and it
becomes infinite with $N$. It is not hard to see that the width of
the barrier also becomes infinite. Thus the process  of \dz \
emission is very suppressed in the large $N$ limit.

There is another possible mode of instability that was pointed out
to us by Maldacena ~\cite{pc}. It is possible for the \ds \ to
nucleate a second spherical \dt \ at a small radius. In this
configuration the strings from the original outer 2-sphere
terminate on the concentric inner 2-sphere. If the system lowers
its energy when the inner sphere expands, it will be unstable, the
inner and outer spheres annihilating one another.

The potential for the inner sphere can be obtained from that of
the outer sphere, eq. (3.14) by making two changes. First of all,
since the \dz \ charge on the inner sphere vanishes $N$ should be
set to zero. Secondly since the strings are now on the outside of
the $2$-sphere the sign of the last term in (3.14) should be
changed. The result is a vanishing potential which indicates that
the inner brane is in neutral equilibrium. Thus there is no
tendency for the inner brane to expand, at least within the
context of our approximations.

Klebanov ~\cite{pc} has pointed out a way to stabilize the inner
brane at vanishing radius. If we retain the full form of the
harmonic function $h$ in eq. (3.2) the perturbation due to the
first term leads to a correction which makes the potential minimum
when the inner brane vanishes.

The stability of the \qhs \ with respect to non-spherically
symmetric perturbations has not yet been carried out.
\begin{figure}
\centerline{\epsfysize=2.00truein \epsfbox{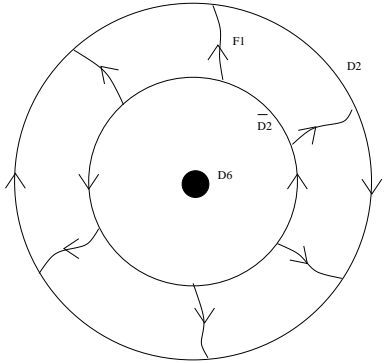}} \vskip
-0.4 cm
 \caption{Two D2-branes surround a D6-brane.
 The two D2-branes have opposite orientation. There are $N$ units
 of flux on the outer D2-brane. $K$ strings now stretch between
 the two D2-branes.
 } \label{diagrams}
\end{figure}

\setcounter{equation}0
\section{The Membrane Theory}

In this section we work in some detail the theory describing the
fluctuations of the membrane. The proper size of the sphere grows
like $N^{1/2}$ in string units. Thus we can focus on a patch much
larger than the magnetic length and approximate it as flat. We
choose co--ordinates  such that the metric is the standard flat
metric. The cartesian coordinates in the \dt \ will be called
$x^i$, $i=1,2$.

Without the $D0$-branes and at low energies, the theory describing
the fluctuations of the $D2$-brane is expected to be a $U(1)$
abelian gauge theory. Now let us dissolve the $N$ $D0$--branes.
Dissolving the $D0$--branes essentially turns the membrane into a
non--commutative membrane. In general, the low energy dynamics of
the theory is expected to be governed by a $U(1)$ non--commutative
Yang Mills theory. However, this is not the end of the story. As
we have seen in eq. (2.4) the presence of the $D6$--branes induces
a $D0$--brane magnetic field ( not to be confused with the world
volume magnetic field $B$ on the \dt \ ). Thus a single \dz \ in
this field will experience a Lorentz force governed by a term in
its Lagrangian
\be
L = {\mu_0 H_3 \over 2} \epsilon_{ij}  X^{i}D_tX^j, \ee where $i,j
= 1,2$
 and
\be
D_tX=\dot X - i[A_0 , X].\ee

For studying the  many \dz \  system we use Matrix Theory
~\cite{BFSS}. The Matrix Theory action corresponding to eq. (5.1)
is
\be
L = {\mu_0 H_3 \over 2} \epsilon_{ij} Tr X^{i}D_tX^j, \ee where
$i,j = 1,2$. This Lagrangian is invariant under the infinitesimal
gauge transformations\be X \rightarrow X + i[\lambda(t), X]. \ee
Indices are raised and lowered by the `closed string' metric
$g_{\mu\nu}$ which we choose to be the standard one.

The effect of this term is two-fold. It first of all induces a
background charge density as in eq. (2.8). In addition it produces
a Chern Simons coupling ~\cite{susskind}. To find the new
couplings we construct a large membrane from $N$ $D0$ branes
moving in a constant background $H$--field, $H_3 = 2\pi \mu_6/V$,
where $V$ is the volume of the membrane.

Following \cite{seiberg}, we  choose matrices $x^1$ and $x^2$ such
that
\be
[x^i , x^j] = i\theta \epsilon^{ij} \ee and set
\be
X^i = x^i + \theta \epsilon^{ij}A_j(x^i). \ee The $x^i$'s are
constant matrices to be identified with the non--commuting
co-ordinates of the membrane. Such matrices exist strictly for
infinite $N$ and are classical solutions to the equations of
motion. The $A_j$'s are fluctuations around the classical
solutions $X^i_{clas}=x^i$ and these will map to the gauge field
living on the brane. Any matrix can be expressed in terms of
finite sums of products $e^{ipx^1}e^{iqx^2}$; so the $N\times N$
matrices $A_i$ can be thought of as functions of the $x^i$'s.

We now insert eq. (5.6) in (5.3) to find an effective Lagrangian
for the fluctuations $A_i$. Dropping total time derivatives we end
up with
\be
L_{eff} = {\pi K \over  V} \left(i\epsilon_{ij}Tr[x^i,x^j]A_0 -
\theta^2\epsilon^{ij}TrA_i\partial_{t}A_j + 2i\theta
\epsilon_{ij}Tr[x^i, \epsilon^{jk}A_k]A_0 + 2i \theta^2
\epsilon^{ij}Tr [A_i, A_j]A_0 \right). \ee Using eq. (5.5) and
\be
[x^i , f]=i\theta\epsilon^{ij}\partial_jf, \ee we can simplify
this as follows
\be
L_{eff} = {\pi K \over  V} \left(-2\theta TrA_0 -
\theta^2\epsilon^{ij}TrA_i\partial_{t}A_j + \theta^2
\epsilon^{ij}Tr\partial_{i}A_jA_0
-\theta^2\epsilon^{ij}Tr\partial_iA_0A_j + 2i \theta^2
\epsilon^{ij}Tr [A_i, A_j]A_0 \right). \ee Finally, introducing
the totally antisymmetric tensor $\epsilon^{\mu\nu\rho}$, we can
write this as
\be
L_{eff} = -{2\pi K \over V}\theta TrA_0 +{\pi K \over  V}\theta^2
\epsilon^{\mu\nu\rho}\left(TrA_{\mu}\partial_{\nu}A_{\rho} +
{2\over 3}iTrA_{\mu}A_{\nu}A_{\rho} \right). \ee

Now we pass to the continuum limit taking $N$ large. We identify
as usual
\be
\theta Tr \leftrightarrow \int {dx^1 dx^2 \over 2\pi}. \ee This
requires that
\be
\theta = {V \over 2\pi N}. \ee We see that $\theta$ is nothing
more than the inverse magnetic field $B$ through the brane. The
magnetic length sets the scale of non--commutativity
~\cite{Connes}\cite{SW}.

The $N\times N$ matrices $A_i$ will map to smooth functions
$A_i(x^i)$ of the non--commutative co--ordinates $x^i$. Since the
fields are functions of non--commuting co--ordinates, we need to
define a suitable ordering for their products in the Lagrangian. A
suitable ordering is Weyl ordering which means that ordinary
products are replaced by the non--commutative $*$ product. In all,
we end up with the following action
\be
S_{eff} = \int d^3x A_0 J^0 +  { K \over 4\pi N}
\epsilon^{\mu\nu\rho}\left(A_{\mu}*\partial_{\nu}A_{\rho} +
{2\over 3}iA_{\mu}*A_{\nu}*A_{\rho} \right). \ee Here, $J^0 =-K/V$
and so there is a net background charge $-K$ on the brane. To
cancel this background charge we add $K$ string ends on the brane.
The action is a $U(1)$ NC CS action at level $K/4\pi N$ and
non--commutativity parameter $\theta = V/2\pi N$ plus the chemical
potential term $A_0J^0$.

In addition to the terms induced by $H$ there is a $U(1)$ NC
Maxwell term. This term has been constructed in \cite{seiberg}. It
is given by
\be
{1 \over 2g_{YM}^2} \int d^3x (det G_{\mu\nu})^{1 \over 2}G^{\mu
\rho}G^{\nu \sigma}\  (F+\Phi)_{\mu \nu} * (F+\Phi)_{\mu \nu} \ee
where $F_{\mu \nu}=\partial_{\mu}A_{\nu}-\partial_{\nu}A_{\mu} +
[A_{\mu},A_{\nu}]_{*}$ and $\Phi_{\mu\nu} =
-\theta^{-1}_{\mu\nu}$. The indices are contracted with the
effective metric $G_{\mu \nu}$
\be
G_{00}=1, \,\ G_{11}=G_{22}= -(2\pi\alpha')^2 \theta^{-2}, \ee and
the coupling constant is given by
\be
g_{YM}^2 \sim {g_s(\rho_*) (2\pi \alpha' \theta^{-1}) \over l_s}.
\ee The dimensionless string coupling constant $g_s(\rho_*)$ is of
course a function of the distance from the $6$--branes. As we
found in eq. (3.19) $g_s(\rho_*) \sim 1/\nu\sqrt{N}$.

As noted by Seiberg \cite{seiberg} the action in eq. (5.14) is of
the usual \nc \ type except for the shift of the field strength by
amount $\Phi$. This shift is of course due to the presence of a
background magnetic field. The Lagrangian differs by the standard
minimal Lagrangian, $F^2$, by a constant term and a total
derivative. Although this makes no change in the equation of
motion, it does have the effect from shifting the value of $F$
from zero to $-\Phi=\theta^{-1}$ in the ground state.

As we found in eq. (3.18) the volume of the membrane (measured in
closed string units) scales like
\be
V \sim N \alpha'. \ee Then
\be
\theta \sim \alpha', \ee and the separation of the constituent
$D0$--branes is fixed in string units. The effective metric is
then
\be
G_{00}=1, \,\ G_{ij} \sim ( -1, -1)\ee and the Maxwell coupling
constant is given by
\be
g_{YM}^2 \sim {g_s(\rho_*) \over l_s} \sim {1 \over \nu
\sqrt{N}l_s}. \ee

Although the Chern-Simons term is interesting, it has nothing to
do with the usual Chern-Simons description of the Quantum Hall
fluid of electrons. This electron fluid may also be described by a
CS theory ~\cite{susskind}. This suggests that the coupled system
of sting ends and $D0$-branes may be described by two coupled CS
theories, one describing the electron fluid and the other the
fluid of \dzs.

The CS term in eq. (5.13) does not influence the physics at scales
much smaller than the size of the entire 2-sphere. This is because
the gauge coupling is very weak. For example the gauge boson mass
induced by the \cs \ term is
\be
m_{ph} \sim 2g_{YM}^2 \nu \sim {1\over \sqrt{N} l_s}. \ee In other
words the Compton wavelength of the photon is of order the sphere
radius. At somewhat shorter distances the forces are dominated by
the ordinary $2+1$ dimensional Coulomb repulsion. At distances
smaller than the string scale the forces are softened by the
effects of \ny \ and other stringy effects. The fact that the
Compton wavelength of the photon is so large means that there is
no meaningful effect on the statistics of the charges, at least
when they are separated by distances smaller than the size of the
sphere. For larger distances the Chern-Simons term may introduce
phases but this should not affect the local physics on smaller
scales.

Thus far we have discussed the gauge field on the \dt . There are
additional world volume fields such as scalars and spinors which
all have similar mass and are described by the appropriate \nc\
fields. However the list of degrees of freedom would not be
complete without the all important electrons. From the point of
view of the \dt \ Matrix Theory, these are not described by
matrices but rather column vectors (or the conjugate row vectors).
Such fields form fundamental representations of the \nc \ gauge
invariance and we describe them by either fermionic or bosonic
fields $|\Psi\rangle$ and $\langle \Psi^{\dag}|$. The appropriate
gauge invariant action for these fields is very simple:
\be
L_{\Psi} =\langle \Psi^{\dag}| (i\partial_t + A_0)|\Psi\rangle +
\langle \Psi^{\dag}|m|\Psi\rangle. \ee The full action is obtained
by adding eq. (5.10) to $L_{\Psi}$
\be
S_{eff} = - {K \over N} TrA_0 + \langle \Psi^{\dag}| (i\partial_t
+ A_0)|\Psi\rangle + \langle \Psi^{\dag}|m|\Psi\rangle + ... . \ee
Varying this action with respect to $A_0$ and taking the trace we
find that the total number of electrons is $K$
\be
\langle \Psi^{\dag}|\Psi\rangle = K. \ee

\setcounter{equation}0
\section{D6-Brane Dynamics}
The near horizon physics of the $D6$-brane system is described by
a $6+1$-dimensional theory which at long distances is a
supersymmetric gauge theory. Indeed the configuration we are
studying may be thought of as a soliton of the $D6$-brane theory.
The only charge carried by the soliton is the $D0$-brane charge
$N$. The spherical $D2$-brane carries no net charge. To interpret
the charge $N$ in the $SU(K)$ gauge theory, we recall that there
is a coupling between the $D6$-brane worldvolume gauge field $F$
and the bulk field $C_1$
\be
\int_7 C_1\wedge F \wedge F \wedge F.\ee Since $C$ is sourced by
the $D0$-brane charge, it follows that our configuration satisfies
\be
\int_6 F\wedge F\wedge F \sim N. \ee Such a classical gauge
configuration is unstable with respect to collapse; that is, it
wants to collapse to zero size. Evidently, this behavior is
resolved in the quantum theory by the $D2$-brane system. Although
the soliton is not absolutely stable, in the limit $N,K
\rightarrow \infty$, the tunneling barrier for the emission of a
$D0$-brane from the $D2$-brane becomes infinite.

Let us consider the strength of the $SU(K)$ couplings on the
$D6$-brane system. The gauge coupling is given by
\be
g_6^2=g_s(\rho). \ee In this formula, $g_6^2$ refers to the
dimensionless coupling at the proper length scale $l_s$. Next we
use
\be
g_s(\rho)={2 \over K}{\left({\rho \over l_s}\right)}^{3 \over
4}.\ee The `t Hooft coupling is given by
\be
Kg_s(\rho)\sim{\left({\rho \over l_s}\right)}^{3 \over 4}.\ee This
equation makes it appear that the coupling vanishes as we approach
$\rho=0$. However, the gauge coupling has dimensions of length to
the cubic power. To determine the effective dimensionless coupling
at a co-ordinate length scale $\Delta \tilde{y}$, we should divide
by three powers of the corresponding proper length. From the
metric, eq. (3.6), we see that the proper length is given by
\be
{\left(\rho \over l_s\right)}^{1 \over 4}\Delta \tilde{y}. \ee
 We need, therefore,  to divide by $(\rho/l_s)^{3/4}{\Delta \tilde{y}}^3$.
so that  the strength of the dimensionless coupling at a
co-ordinate scale $\Delta \tilde{y}$ is given by
\be
{\left(l_s^3 \over \Delta \tilde{y}\right)}^3. \ee
Thus at $\Delta
\tilde{y}$ of order one in string units, the $D6$-brane theory
becomes strongly coupled.

Now consider the $K$ string ends on the $D6$-brane. These objects
are analogous to non-relativistic quarks in QCD. Their gauge
interactions become strong at separations $\Delta \tilde{y} \sim
l_s$. Let us assume that they bind into an $SU(K)$ singlet,
``baryon,'' of this size. We would like to compare the energy
scales of the baryon-excitations with the energy scales discussed
in section (4). In this discussion, energy means conjugate to
$\tau$.

The excitation energy of the baryon is of order one in string
units since the natural scale is $\Delta \tilde{y} \sim l_s$. As
we saw in section (4), the proper energy ($T$-energy) of string
oscillations, higher Landau levels and quasiparticles is of order
$N^{-1/2}$ in string units. To convert this to $\tau$-energy, we
need to multiply by a factor ${g_{00}}^{1/2}$ at the $D2$-brane.
For example, the quasiparticle $\tau$-energy is given by
\be
E_{Quas}\sim {{\rho_*}^{1 \over 4} \over \sqrt{N}} \sim N^{-{1
\over 3}}.\ee The implication is that the energy scale of the
baryon-excitations is much larger than the excitation scales of
the $D2$-brane. In the sense of the Born-Oppenheimer method, the
baryon degrees of freedom are fast degrees of freedom.

\setcounter{equation}0
\section{Properties of the Electron System}
The string-ends that move on the $D2$-brane are charged particles
with respect to the membrane world-volume gauge theory. We will
refer to them as electrons. In this section we discuss their
properties.

\subsection{The Statistics of the Charges}
As we will see, the question of the statistics of the electron
string-ends on the $D2$-brane is far from straightforward.

Consider a ground string state connecting the $D6$ and $D2$
branes. General string theory arguments given in the appendix tell
us that these strings satisfy fermionic statistics. However, the
fact that the full $6-2$ strings are fermions, does not imply that
the electrons on the $D2$-brane are fermions. A simplified model
illustrates the subtleties. We will assume that the $6-2$ strings
remain in their ground state apart from the motion of their
end-points on the branes. In our approximation, the motion of each
of the two string ends is independent of the motion of the other.
Then a string is characterized by a location on the $D2$-brane $x$
and a location on the $D6$-branes $y$. In addition, it has an
$SU(K)$ index $i$ labeling which $D6$-brane it ends on. As we saw,
the strings are fermions. The $K$-body wavefunction has to be
antisymmetric with respect to simultaneous interchange of any pair
of labels $(x,y,i)$.

Concerning the $SU(K)$ indices we  assume that they combine to
form a singlet. In the previous section, we discussed the gauge
interactions on the $D6$-brane. Although, the string coupling
vanishes at the $D6$-brane, the `t Hooft coupling is large at
length scales $\Delta \tilde{y} \sim l_s$. Therefore, we expect
that any non-singlet configuration would radiate gauge bosons
until it discharged. Since the singlet wavefunction is
antisymmetric with respect to the $SU(K)$ indices, the remaining
wavefunction must be symmetric.

The most naive assumption about the behavior of the ends on the
\ds \ is that they are  all localized at $y=0$. This would mean
that the wavefunction is symmetric with respect to interchange of
the $y$ coordinates. In this case the electrons are bosons since
the dependence on the $x$ coordinates is also symmetric.

The reason that this may be naive is that the gauge forces on the
\ds \ between string ends may not be weak if they are localized
with small separation. In other words the dynamics of the ``knot"
where all the strings come together may be non-trivial. Perhaps it
is possible that an antisymmetric sector for the baryon
wavefunction exists.

We will consider two possible sectors of the theory. In the first
sector the wavefunction of the $y$ co-ordinates is symmetric and
the electrons are bosons while in the second sector it is
antisymmetric and the electrons are fermions. We do not know which
sector has the lower energy but as far as the fast dynamics of the
$D6$-brane interactions is concerned, the two sectors are
uncoupled superselection sectors.  Thus
the $K$-body wavefunction of the strings can have either one of
the following forms
\be
\Psi_1=\phi_s(x_1,...,x_k)\epsilon_{i_1i_2..i_k}{\psi_s}^{i_1i_2..i_k}(y_1,...,y_k),\ee
and
\be
\Psi_2=\phi_a(x_1,...,x_k)\epsilon_{i_1i_2..i_k}{\psi_a}^{i_1i_2..i_k}(y_1,...,y_k),\ee
where $\phi_s,\psi_s$ and $\phi_a,\psi_a$ are symmetric and
antisymmetric functions of their arguments respectively.

Our primary interest in this paper is in the physics described by
the wavefunctions $\phi(x)$. These wavefunctions describe the
physics of $K$ charged fermions, if $\phi$ is antisymmetric, or
$K$ charged bosons if $\phi$ is symmetric. The particles move on a
$2$-sphere with $N$ units of magnetic flux. Thus the low lying
spectrum of states should be that of the bosonic or fermionic
Quantum Hall system with filling fraction
\be
\nu = {K \over N}. \ee Without further evidence we will assume
that the conventional Quantum Hall phenomenology applies to our
system. For example, we assume incompressible Quantum Hall states
exist for all odd denominator $\nu$'s in the fermion case and even
denominator $\nu$'s in the bosonic case.

\subsection{Quasiparticles}
An important feature of the Quantum Hall effect is the existence
of an energy gap and fractionally charged quasiparticles.
Let us briefly review the
construction of these objects ~\cite{Laughlin}\cite{anyon}.

First begin with the theory on the plane. The lowest Landau level
(LLL) wavefunctions are degenerate and there is one orthogonal LLL
for each unit of magnetic flux. It is helpful to make an
identification of the LLL's with the flux quanta. In the stringy
construction in this paper, the flux quanta are the $D0$-branes.
Each $D0$-brane can be thought of as a LLL and a string ending on
that $D0$-brane is an electron in that LLL. Since the $N$ LLL's
are degenerate, there is a $U(N)$ symmetry of the space of LLL's.
This $U(N)$ symmetry is just the $U(N)$ gauge invariance of the
Matrix Theory description of $D0$-branes. It is also a regularized
version of the area preserving diffeomorphism group.

The conventional construction of a quasiparticle begins with the
idea of an infinitely thin solenoid passing through the substrate
\cite{Laughlin}. The magnetic field through the substrate is
adiabadically increased until the flux equals one Dirac unit. The
new gauge field is a gauge transformation of the old, but the
process induces a change in the state of the system. To understand
the change, it is convenient to work in a basis of angular
momentum LLL's, $|l>$. The individual angular momentum
wavefunctions are concentrated on circular rings of radius $\sim
l^{1/2}$ with the solenoid at the center. Turning on the
solenoid-flux, takes each electron in the $l$-th state to the
$l+1$ state but in the process the $l=0$ state is left unoccupied.
The result is a hole in the electron density. Since each LLL had
originally an average charge $\nu=K/N$, the hole has charge $\nu$.
The radius of the hole is just the magnetic length and it is
independent of the charge.

Another way to construct the quasiparticle is to begin with a
distant magnetic monopole one one side of the substrate.
Adiabadically passing the magnetic monopole through the substrate
to the other side has the same effect as turning up the current in
the solenoid. The monopole picture is especially relevant for the
spherical substrate. Transporting the monopole from outside to
inside the sphere creates an additional unit of magnetic flux but
does not increase the number of electrons. The result is a hole at
the place where the monopole passed through the sphere.

An intuitive way to think about these effects is to picture the
magnetic flux as an incompressible fluid with the electrons moving
with the fluid. When a new unit of flux is added it pushes the
fluid away, creating a hole in the electron density. As we have
seen the \dz \ fluid does in fact behave incompressibly.

In the string/brane setup of this paper, there are neither
solenoids nor monopoles. In fact, the $2$-sphere does not divide
space into an inside and outside. A possibility that comes to mind
is to pass a $D6$-brane through the $2$-sphere but this has the
effect of changing the electron number by one unit, not the flux.

The key to the formation of the quasiparticle is the $D0$-brane.
We have previously seen that the dissolved \dzs \ form an
incompressible fluid. We introduce an additional $D0$-brane far
away from the substrate $2$-sphere. Now adiabadically allow it to
approach the $2$-sphere at some point $x_0$.  At some distance of
order $l_s$, it will get absorbed by the $D2$-brane adding a unit
of flux at $x_0$ to the original $N$ units. The flux behaving like
an incompressible fluid will increase the area of the sphere by
one unit, leaving a hole of charge $\nu$ in the charged particle
distribution at the point $x_0$.

The quasi particle defined in this way is not necessarily stable.
As an example consider the fermionic case $\nu = {p /(2p+1)}$ with
integer $p$. Now take two quasiparticles of charge $\nu$ and
combine them with one extra electron. If they bind, the result is
a new quasiparticle of charge $-1/(2p+1)$. (It is also possible to
create an excitation with charge $+1/(2p+1)$). In this case the
original quasiparticle can decay into $p$ constituents. The \qp \
with charge $-1/(2p+1)$ can be constructed by starting with an extra
$6-2$ string with two \dzs\ attached to it. By sliding the \dzs\
toward the membrane until they dissolve, the new \qp \ is created.
In the limit $p \to \infty$ the neutral \qp \ of the $\nu = 1/2$
state results \cite{Reed}.

\begin{figure}
\centerline{\epsfysize=2.00truein \epsfbox{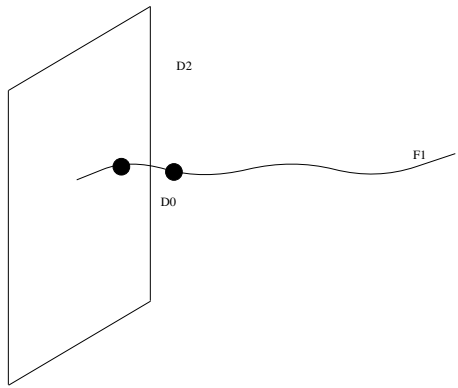}} \vskip -0.4
cm
 \caption{The figure shows two D0-branes on a fundamental string.
 The string ends on a D2-brane.}
 \label{diagrams}
\end{figure}

\subsection{Composite Fermions}
The qualitative features of the QHE have been nicely captured by a
phenomenological model, the so called composite fermion model
(CFM) ~\cite{Jain}. While the theoretical underpinnings of the CFM
are not completely secure, it does appear to successfully
correlate many properties of the various fractional QHE ground
states.

We make no claim in this paper to deriving the CFM. However we do
think the language of string theory is suggestive and might offer
new insights. We regard it as a challenge to use the tools of
string theory and \nc \ field theory to give a derivation. What we
will do is to explain how to state the rules of CFM in terms of
string theoretic concepts. We will assume that the electrons are
fermions although the arguments are easily generalized to the
bosonic case.

Up to now we have thought of the magnetic field as representing
the density of \dzs . Now we want to change perspective a bit.
Recall that in string theory there is a gauge invariance
associated with the $NS-NS$ 2-form potential $B_{\m \nu}$. The
magnetic field $F$ on the \dt \  is not gauge invariant but should
be replaced by $B+F$. The electrons feel the field $B$ as a
background magnetic field. According to the new perspective we
will consider $B$ to be a background magnetic field and $F$ to be
the density of \dzs . In other words some fraction of the \dz \
density can be replaced by the background $B$ field. This leaves
over a number of \dzs \ $N'= N-\int B$. For example let us take
$N' = K$. In this case we have divided the field into background
and \dzs \ in such a way that there is exactly one \dz \ for each
string end. In this picture each string ends on a unique \dz . The
basic assumption of the CFM is that we may think of each string
end as bound to a \dz \ forming a composite.

The idea that the \dzs \ move with the string ends may have more
to do with the gauge invariance of Matrix Theory than with any
dynamical attraction between the string-ends and the \dzs . Let us
think of the string-ends as distinguishable particles but with the
wave function being appropriately symmetrized at the end. We can
label the strings from $1$ to $K$. If we  choose $N'=K$  there is
exactly one \dz \ for each string. Recall that the labeling of the
\dzs \ in Matrix Theory is a choice of gauge in the Super Yang
Mills quantum mechanics. There is a particular choice of gauge in
which the $nth$ string is defined to end on the $nth$ \dz . In
this gauge each string is attached to a specific \dz .

A second assumption is that when bound to an string-end a \dz \
acts as a fermion so that the composite has opposite statistics
from the original string end. This assumption can be motivated
from the fact that the a \dz \ behaves like a unit of flux.

Putting these assumptions together  we conclude that the $K$
electron system in $N$ units of flux can be replaced by a system
of  of $K$ opposite-statistics electrons in $N-K$ units of flux.
Thus for example, the $\nu =1$ fermionic system is equivalent to a
system of bosons in no field. Similarly  the $\nu=1$ boson system
is a free fermion system. Repeated use of these rules generates
the full CFM.

\setcounter{equation}0
\section{Modeling Edges}

Some of the most interesting phenomena in the Quantum Hall system
are associated with the edges of the sample. To model the edges we
can modify the system by introducing a single NS 5-brane into the
system, thus providing a boundary for the membrane.

The 5-brane is oriented in the $X^1,X^2,Y^1,Y^2, Y^3$ directions
and is located at the origin of the other coordinates. It
intersects the 2-sphere on the equator \bea X^3 &=&0 \cr (X^1)^2 +
(X^2)^2 &=& r^2. \eea

The NS 5-brane intersects the \ds \ forming a stable BPS
configuration. The intersection of the \dt \ and the  NS 5-brane
is also stable for large radius. In this case the sphere is almost
flat and the membrane intersects the 5-brane orthogonally.
Furthermore the 5-brane acts as a boundary for the \dt \ and
allows us to consider only the hemisphere $X^3\geq 0$.

There is a subtlety concerning the \dzs \ in this case. A zero
brane can be bound in the 5-brane as well as in a \dt . In fact
one can expect a \dz \ to escape from the \dt \ into the 5-brane.
Since the 5-brane is infinite the \dz \ will escape to infinity
along the 5- brane.  The way to prevent this is to fill the
5-brane with a constant \dz \ charge density. By choosing this
density large enough we can insure a net charge on the \dt .
Another way to think of this is to imagine boosting the
intersecting 2 and 5 branes along the 11th direction of M-theory.
The momentum will be shared between the branes in a way which is
controlled by requiring their velocities to match.  As in the case
without the $NS5$-brane, the \ds \ continues to repel the \dz \
and leads to an equilibrium as before. The single 5-brane is a
small perturbation on the metric of the $K$ \dss .

To understand the effect of the 5-brane on the electrons, recall
that a string can not end on an $NS5$-brane. Accordingly, the
5-brane is a repulsive ``brick wall" to the electrons. To estimate
the effect of this brick wall we consider the quantum mechanics of
a non-relativistic charge in a magnetic field in the presence of a
brick wall, in other words on a half-plane. In an appropriate
gauge the Hamiltonian is
\be
H={1\over 2m}\left[ (p_x -eBy )^2 +p_y^2
 \right].\ee By diagonalizing $p_x$ and shifting the origin of $y$ we
obtain an harmonic oscillator for each value of $p_x$. The ground
state of this oscillator is the LLL for that value of $p_x$.

The effect of the 5-brane is to force the wave function to vanish
at $y=0$.

Let us begin with the state with $p_x =0$. The Hamiltonian is a
conventional oscillator in this case. The relevant sector of the
oscillator is the states with odd wave functions which vanish at
$y=0$. Thus the ground state is described by a wave function of
the form
\be
\sqrt{eB}y \exp {({-\12}eBy^2)}. \ee

Now let us consider the effect of $p_x \neq 0$. We will do this in
perturbation theory. From eq. (8.2) we find the lowest order
perturbation to be
\be
\delta H= -{eB p_x \over m}y. \ee We find the leading dependence
of the  energy on $p_x$ to be
\be
\delta E = \sqrt{eB} {p_x \over m}. \ee

For the theory on the spherical \dt \ $eB \sim 1$ and $m \sim
N^{1/ 2}$ in string units. Thus the energy of a string-end near
the $NS5$-brane is given by
\be
E = N^{- 1/2} p_x. \ee These modes behave like right-moving
massless excitations moving with fixed velocity $\sim N^{-1/2}$ on
the boundary of the hemisphere. These are the expected edge
states. Note that the energy gap associated with these states is
obtained when $p_x $ takes its minimum value $\sim N^{-1/2}$. Thus
the gap is of order $1/N$ in string units. This is parametrically
smaller than the typical energy scale $N^{- 1/2}$ as it should be.

The physics of the fluid of string-ends near the  $5$-brane is
complicated but it should be described by a $1+1$ dimensional
conformal field theory. We do not know how to derive this field
theory from the underlying string theory but the phenomenolgy of
the quantum hall effect suggests that it is described as a
Luttinger liquid  with excitations carrying the same statistics as
the bulk quasiparticles. For filling (fermionic) fractions $\nu
=1/(2p+1)$ these quasiparticles have statistics equal to $\nu$.
This last point may be somewhat complicated due to the
Maxwell-Chern-Simons term in the action (5.23) which introduces
phases when a quasiparticle moves relative to a second
quasiparticle at distances of order the radius of the sphere.

\begin{figure}
\centerline{\epsfysize=2.00truein \epsfbox{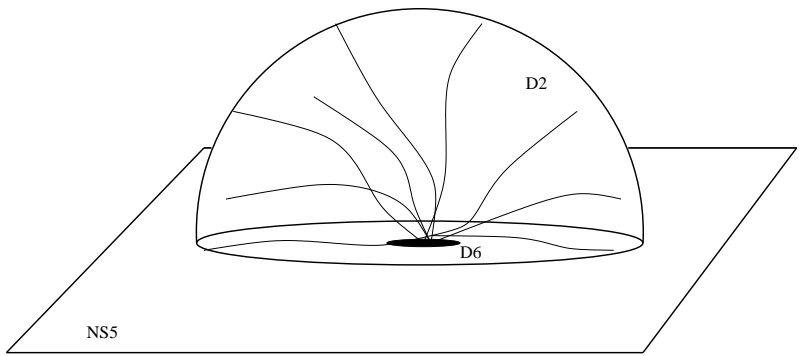}} \vskip -0.4
cm
 \caption{Here we see a hemi-spherical D2-brane ending on an NS 5-brane.
 $K$ D6-branes and embedded in the NS 5-brane. $K$ strings stretch to the
D2-brane. }
 \label{diagrams}
\end{figure}

\setcounter{equation}{0}
\section{Giant Gravitons}

One may wonder whether the \qhs \ has an eleven dimensional
meaning. Up till now we have treated the coupling constant as if
it tended to a finite value at infinity. However the actual value
of the asymptotic coupling cancels from all of our results
concerning the behavior of the $D2$-brane. This allows us to take
a limit in which the asymptotic coupling $g_s$ tends to infinity.
This is the limit in which the 11th dimension decompactifies.

Thus the Quantum Hall solitons are naturally thought of as objects
in eleven dimensions. First consider the \ds . The
$11$-dimensional origin of the \ds \ is itself not a brane but a
Kaluza Klein monopole consisting of a product of $7$-dimensional
Minkowski space and a $4$-dimensional Taub-Nut space with mass
parameter $K$.

Now add a graviton to this space and boost it so that its momentum
along the $11$th direction is $N$ in quantized units. In the IIA
description this is a configuration with a \ds \ and $N$ units of
\dz \ charge.  It has exactly the quantum numbers of the Quantum
Hall soliton.

Ordinarily the Taub-Nut monopole will repel the graviton and send
it off to transverse infinity. However our results indicate that
the graviton can be trapped or bound to the center of the monopole
where it will have the rich excitation spectrum of the Quantum
Hall soliton.

From the results of section (2) we see that the graviton grows
with increasing momentum. In this sense the soliton is similar to
the giant gravitons of Ref. ~\cite{Mcgreevy} with the background
n-form field ~\cite{Myers} being replaced by the Taub-Nut
background.

The $11$-dimensional interpretation makes the most sense if we fix
$K$ and let $N$ grow. This corresponds to boosting the graviton in
a fixed background. In this case we are really discussing a fixed
number of charges in the background magnetic field.

\setcounter{equation}{0}
\section{Conclusions}
In this paper we have constructed a system of branes and strings
whose low energy excitations are described in terms of
non-relativistic particles moving on a $2$-sphere in a magnetic
field with repulsive gauge forces between the particles. We have
found that there is a characteristic energy scale for the low
energy excitations and that all energies associated with the two
dimensional non-relativistic system are of this scale. Thus we
have all the ingredients for a string theory simulation of the
Quantum Hall system.

The background magnetic field may  be described in terms of a
density of \dzs \ dissolved in the \dt \ substrate. The \dz \
fluid behaves like an incompressible fluid. The \dzs \ play the
role of quantized units of flux. In this picture quasiparticles of
the QH system are simply additional \dzs .

Alternatively the field may be described in terms of a background
$2$-form $B_{\mu\nu}$ field. More generally by choosing a gauge,
the field can be represented as a combination of \dzs \  and $B$
flux. We argued that this gauge freedom is closely connected with
the so called Composite Fermion Model of the fractional Quantum
Hall Effect. Seiberg ~\cite{pc} has suggested that this freedom of
description may be related to the ambiguity in the definition of
the $\Phi$ parameter in eq. (5.14) ~\cite{seiberg}.

We briefly discussed the modeling of edges and edge states in the
system by introducing an $NS5$-brane. The $5$-brane intersects the
spherical membrane along its equator and produces a boundary along
with the typical chiral edge states.

A dual way of looking at the \qhs \ is in terms of the near
horizon gauge theory of a stack of \dss . From this point of view
the configuration is a metastable soliton of the theory carrying
$F\wedge F\wedge F$ charge. The existence of the soliton with its
very rich spectrum of low energy excited states is new information
about the $6$-brane.

A final point concerns compactification. In this paper we have
considered the case of uncompactified $IIA$ string theory.
However, we see no obstruction to compactifying the six dimensions
$Y^{a}$. In this case, our configurations would exist as
metastable objects in the $3+1$-dimensional world.

Needless to say, we hope that string theory techniques will be
useful in understanding the Quantum Hall system and other
condensed matter systems and conversely, that condensed matter
phenomenology may teach us new lessons about string theory.

\setcounter{equation}0
\section{Appendix}
Consider a ground string state connecting the $D6$ and $D2$
branes. General string theory arguments tell us that these strings
satisfy fermionic statistics. To see this, we begin with a brane
configuration in which the $D6$-brane is oriented along the $Y^a$
directions, $a=6,...,9$ and a $D2$-brane along the $X^i$
directions, $i=1,2$. The two branes may be separated along the $3$
direction.

Let us recall why the string ground state is a fermion. We are
interested in the spectrum of the $6-2$ strings. The symmetry of
the problem is $SO(2) \times SO(6)$. The total number of
worldsheet fields which satisfy mixed boundary conditions is
eight: $X^{1,2}$ satisfy $DN$ boundary conditions while the
$Y^{a}$ satisfy $ND$ boundary conditions. $X^3$ satisfies $DD$
boundary conditions. The boundary conditions break the Lorentz
symmetry in this problem.

As usual, in the $R$ sector, bosonic and fermionic fields satisfy
the same periodicity conditions and the zero point energy
vanishes. Therefore, in the $R$ sector the ground states are
massless. The only fermionic worldsheet field that are periodic
are $\Psi^0$ and $\Psi^3$. From these we get zero modes, and,
therefore, an extra degenerate ground state. The two states have
opposite worldsheet fermion numbers. Thus only one ground state
survives the GSO projection. The surviving ground state is a
singlet under the symmetry group $SO(2) \times SO(6)$.

The fact that the Ramond ground states are fermions can be deduced
as follows. Let us do three $T$-dualities along the
$1,2,3$-directions to turn the system into a $D1$-$D9$ brane
system. $T$-duality is a gauge symmetry of the theory and should
not change the spectrum or the statistics of the string states.
Now let us look at the $R$ sector of the $1-9$ strings. The
symmetry of the $T$-dual configuration is $SO(1,1)\times SO(8)$
and we can take advantage of this maximal symmetry. The two $R$
ground states are left and right moving spinors of $SO(1,1)$
respectively and singlets under the internal $SO(8)$. Only one of
them survives the GSO projection. The fact that they are fermions
follows from the spin--statistics theorem.

In the $NS$ sector, the zero point energy can be computed in the
usual way giving
\be
-{1 \over 2}+{\# ND \over 8} = {1 \over 2}. \ee Thus the $NS$
sector is massive and therefore the bosons are massive. Separating
the branes along the $3$ direction shifts the overall spectrum of
the $6-2$ strings by a term proportional to the length of the
stretched strings.

Finally, we remark that this configuration leaves two
supersymmetries unbroken. The ground state of the $6-2$ strings is
a BPS multiplet and it consists of a single state with no
bose--fermi degeneracy. This state is fermionic as we have argued
above. Excited states are in long multiplets and these contain
equal numbers of fermions and bosons. The supersymmetry of the
problem is broken by the addition of the $D0$--branes.

For the large spherical brane of section (3), we focus on a patch
along the $1,2$ directions much larger than the magnetic length
and approximate it as flat. Then, we can use the above analysis to
estimate the free spectrum and also the statistics of the string
ground state follows. As long as we focus on charged particles
separated at distances of order the magnetic length, the Chern
Simons term that we found in the previous section does not play
any important role in the statistics.

\section{Acknowledgements}
L.S. would like to thank I. Klebanov, J. Maldacena, G. Moore and
N. Seiberg for useful conversations. J.B. would like to thank U.
Chicago for their hospitality while this work was being completed
and D. Kutasov for stimulating discussions. We also acknowledge M.
Fabinger, B. Freivogel, M. Kleban, J. Mcgreevy, M. Rozali, S.
Shenker, E. Silverstein and S. Zhang for useful discussions. We
thank A. Matusis for collaboration during the early stages of this
work. The work of L.S. and N.T. was supported in part by NSF grant
980115. J.B. is supported under D.O.E. grant number
AE-AC03-76SF-00515.

\end{document}